\documentclass[aps,prb,showpacs]{revtex4}

\usepackage{graphicx}
\usepackage{amsmath}
\usepackage{amsfonts}
\usepackage{amssymb}

\graphicspath{{./IMAGES/}}

\begin{document}
\title{Optimal traps in graphene}

\author{C. A. Downing}
\email[]{c.a.downing@exeter.ac.uk}
\affiliation{School of Physics, University of Exeter, Stocker Road,
Exeter EX4 4QL, United Kingdom}

\author{A. R. Pearce}
\affiliation{School of Physics, University of Exeter, Stocker Road,
Exeter EX4 4QL, United Kingdom}

\author{R. J. Churchill}
\affiliation{School of Physics, University of Exeter, Stocker Road,
Exeter EX4 4QL, United Kingdom}

\author{M. E. Portnoi}
\email[]{m.e.portnoi@exeter.ac.uk} \affiliation{School of Physics,
University of Exeter, Stocker Road, Exeter EX4 4QL, United Kingdom}
\affiliation{International Institute of Physics, Universidade Federal do Rio Grande do Norte, 59012-970 Natal - RN, Brazil}

\date{\today}

\begin{abstract}
We transform the two-dimensional Dirac-Weyl equation, which governs the charge carriers in graphene, into a non-linear first-order differential equation for scattering phase shift, using the so-called  variable phase method. This allows us to utilize the Levinson Theorem to find zero-energy bound states created electrostatically in realistic structures. These confined states are formed at critical potential strengths, which leads to us posit the use of `optimal traps' to combat the chiral tunneling found in graphene, which could be explored experimentally with an artificial network of point charges held above the graphene layer. We also discuss scattering on these states and find the $m=0$ states create a dominant peak in scattering cross-section as energy tends towards the Dirac point energy, suggesting a dominant contribution to resistivity.
\end{abstract}

\maketitle

\section{\label{intro}Introduction}

The electronic properties of the two-dimensional (2-D) material graphene\cite{Origin, Abergel} are of great interest due to the quasi-relativistic nature of its spectrum. Interesting transport effects such as chiral (Klein) tunneling,\cite{Klein, Katsnelson, Hewageegana, Young, Reijnders} vacuum polarization,\cite{Shytov} atomic collapse\cite{ShytovCollapse, DowningCollapse} and the minimum conductivity at the Dirac point\cite{Novoselov} have been widely discussed in the literature. The topic of elastic scattering in clean, low-temperature graphene, which can occur due to charged impurities, ripples or strain fields, has been addressed by many authors.\cite{Adam, Peres, DasSarma}

However, despite its extraordinary properties there is a major obstacle stopping graphene from usurping silicon in the electronics industry, namely is its lack of a bandgap. This frustrates attempts to perform digital logic with graphene due to the difficulty in turning off the flow of chiral charge carriers which always wish to conduct. Attempts at opening a gap in monolayer graphene have focused on cutting into nanoribbons,\cite{Brey} chemical functionalization\cite{Withers} and strain engineering,\cite{ZNi} which can unfortunately blunt the remarkable electronic properties which makes graphene so attractive in the first place. Here we propose a method not to open a gap, but to switch off the chiral tunneling by considering zero-energy states, when the Fermi energy coincidences with the Dirac points, such that fully confined states are predicted to be able to form due to the absence of pseudospin.\cite{Bardarson, Hartmann} In fact, these states are the most important factor when considering resonant scattering in graphene.

Despite the appearance of sophisticated experimental techniques for probing resonances and the modification of the density of states in the continuum\cite{Crommie} the search for fully-confined (square-integrable) states remains a significant ongoing task. Efficient manipulation of the Fermi level requires the presence of a back-gate in close proximity to the graphene, which makes the numerous beautiful results stemming from the long-range behavior of the bare Coulomb potential\cite{ShytovCollapse} to be of somewhat far from experimental reality, as the presence of image charges in the gate material (or screening effects) make any realistic potential fall at large distances faster than $1/r$.\cite{Asgari} It is important to emphasize that any fast-decaying potential cannot produce a bound state at nonzero energy.\cite{Chaplik} Indeed, the asymptotic of the wavefunction is a Bessel function decaying asymptotically only like $r^{-1/2}$, and so one is led to consider zero-energy states instead.    

Whilst low-energy resonant scattering in monolayer graphene has been intensively studied by previous authors both theoretically\cite{Chaplik, Novikov, ShytovCollapse, AndreGeim, Guinea, Basko, Titov, Ostrovsky, Wehling, Schneider} and experimentally\cite{Ponomarenko, Chen, Ni, Monteverde} the importance of fully-confined zero-energy states in realistic structures has not been fully appreciated until recently.\cite{Cresti} Thus far, only quasi-bound states, where only one wavefunction component is confined or when the wavefunction is non-square integrable, have been considered for resonant scattering. Previously, only circular wells\cite{AndreGeim, Chaplik} or the Coulomb potential,\cite{Novikov, ShytovCollapse} have been investigated, but in this paper we concentrate on smooth,\cite{Downing, Kusmartsev}  short-range potentials which are defined with two parameters, characterizing both strength and spread.

We study confined states and resonant scattering in graphene due to either scanning probe microscopy (SPM)\cite{Deshpande} tip-induced potentials or due to some charge displaced out of the plane,\cite{Fogler} with careful consideration of truly bound zero-modes. The strength parameter can arise due to, for example, the size of the charge on the SPM, whilst the spread parameter is linked to the distance from the SPM tip to the graphene. We investigate both the conditions required for a zero-energy bound state to form and the effect of such states in our study of the energy dependence of scattering cross-section and resistivity contributions of resonant scatterers. We also note there is increasing interest in zero-modes of the Dirac equation in the condensed matter community due to the possibilities of both observing the elusive Majorana fermions\cite{Jackiw} or indeed fractionally charged excitations.\cite{Mudry} 

To carry out our investigations into realistic, short-range (due to the necessity of a gate in all measurements) potentials we develop the variable phase method or VPM,\cite{Calogero,Babikov} which was found to be useful for tackling scattering problems governed by the Schr\"{o}dinger equation in 2-D,\cite{Portnoi1988, PortnoiVPM2D, Portnoi1998, PortnoiIonisation, Ouerdane08} for use with the 2-D Dirac-Weyl equation - allowing us to consider the charge carriers of graphene. The VPM was originally developed\cite{MorseAllis} for use with non-relativistic wave equations in the 1930s as a neat tool to calculate physically relevant quantities directly, rather than having to extract them from the wavefunction, and has recently been developed for use with the Dirac-Weyl equation in quasi-one-dimensional problems due to the intense interest in graphene in the condensed matter community.\cite{Stone} Here we derive a first-order equation from which we can immediately find the scattering phase shift. This is advantageous as important physical properties directly follow, such as: the number of bound states (from the Levinson Theorem);\cite{Levinson, Portnoi1998} the scattering and transport cross-sections (from standard elastic scattering theory);\cite{Newton} the number of states around a potential barrier (using the Friedel sum rule);\cite{Friedel} and the energy change due to the impurity interacting with neighboring electrons (the Fumi theorem).\cite{Fumi} We have neglected effects due to rippling of or dislocations in the graphene membrane. We also do not discuss scattering by multiple electrostatic barriers\cite{DowningHeun, DowningWhitt} or by magnetic barriers,\cite{Egger, Zazunov, Ramezani} but our method can be generalized to account for the presence of vector potentials. 

The rest of this work is as follows. We outline the logic behind the VPM and give a detailed derivation of the phase equation for the Dirac-Weyl equation in Sec.~\ref{formalism}. In Sec.~\ref{results}, we compare the results of the VPM against tight-binding calculations on a finite flake and then apply elastic scattering theory in conjunction with the VPM, which allows us to investigate both the nature of confined states in graphene and the influence of short-range scattering on some of its transport properties. Finally, we summarize and discuss our results in Sec.~\ref{disc}. In Appendix~\ref{appendA} we show how quantum confinement of zero-energy states can occur for a quite general class of potential wells decaying like a power law, whilst in Appendix~\ref{appendB} we detail the key results for the simple case of scattering by a circular finite potential well.

\section{\label{formalism}Formalism}
The 2-D Dirac-Weyl Hamiltonian governing the low-energy charge carriers of graphene on a Dirac cone is\cite{Abergel}
\begin{equation} 
\label{DiracEq}
\hat H = v_{\mathrm F} \boldsymbol \sigma \cdot \boldsymbol{\hat p} + U(r) + \sigma_z \Lambda,
\end{equation}
where $v_{\mathrm F}\approx c/300$ is the Fermi velocity of the Dirac particles, $\boldsymbol \sigma=(\sigma_x,\sigma_y,\sigma_z)$ are the Pauli spin matrices, $U(r)$ is a central potential and  we include $\Lambda$ as the mass term for generality. A non-zero mass can arise due to chemical modifications or by strain engineering.\cite{Rozhkov} We move into polar coordinates $(r, \theta)$ for circular symmetry, and separate the variables via the following ansatz for the two-component spinor wavefunction
\begin{equation}
\label{CompleteWF}
\Psi(r,\theta) = \frac{e^{im\theta}}{\sqrt{2\pi}} \left(
 \begin{array}{c}
 \chi_A(r) \\ ie^{i\theta}\chi_B(r)
 \end{array}
\right), \qquad m=0,\pm1, \pm2, ...
\end{equation}
where the subscripts $A$ and $B$ label the two sublattices of the graphene chicken wire lattice. This choice of wavefunction leads to two coupled first-order differential equations for the radial wavefunction components $\chi_{A,B}(r)$
\begin{subequations}
\label{coupledchi}
 \begin{align}
  \left(\frac{d}{dr} + \frac{m+1}{r} \right) \chi_B &= (k-V(r)-\Delta)\chi_A, \label{coupledupper} \\
  \left(-\frac{d}{dr} + \frac{m}{r}   \right) \chi_A &= (k-V(r)+\Delta)\chi_B, \label{coupledlower}
 \end{align}
\end{subequations}
with $V(r)=U(r)/\hbar v_{\mathrm{F}}$, $\Delta=\Lambda/\hbar v_{\mathrm{F}}$ and $k=E/\hbar v_{\mathrm{F}}$, where $E$ is the eigenenergy. Re-arranging Eqs.~\eqref{coupledchi} into a second-order differential equation for a single radial wavefunction component $\chi_{A}(r)$ we obtain for the massless case
\begin{align}
\label{decoupledchi}
 \frac{d^2}{dr^2}\chi_A(r) +\left(\frac{1}{r} +\frac{1}{k-V(r)}\frac{dV(r)}{dr} \right) \frac{d}{dr}\chi_A(r) + \left( (k-V(r))^2 - \frac{m}{r}\frac{1}{k-V(r)}\frac{dV(r)}{dr} - \frac{m^2}{r^2} \right) \chi_A(r) = 0.
\end{align}
We consider potentials of the form $V(r\to\infty)=0$ such that at large distances Eq.~\eqref{decoupledchi} reduces to\cite{note}
\begin{equation}
\label{decoupledchifree}
 \frac{d^2}{dr^2}\chi_A(r) +\left(\frac{1}{r} \right) \frac{d}{dr}\chi_A(r) + \left(k^2  - \frac{m^2}{r^2} \right) \chi_A(r) = 0,
\end{equation}
which is the Bessel equation with the well-known solution $a_{m} J_{m}(k r) - b_{m} N_{m}(k r)$, or equivalently
\begin{equation}
\label{wavefunc}
 \chi_{A}(r) = A_{m}\left[ J_{m}(k r)\cos(\delta_{m}) - N_{m}(k r)\sin(\delta_{m}) \right],
\end{equation}
where $J_{m}(k r)$ and $N_{m}(k r)$ are the Bessel functions of the first and second kinds respectively, and $\delta_{m} = \arctan(b_{m} / a_{m} )$ is the scattering phase shift, arising from the difference in phase of the wavefunction at $r\to\infty$ compared to the free particle case.

We now implement the VPM by treating the constants $A_{m}$ and $\delta_{m}$ as functions of the radial coordinate $r$, such that
\begin{equation}
\label{wavefuncansatz}
 \chi_{A}(r) = A_{m}(r)\left[ J_{m}(k r)\cos(\delta_{m}(r)) - N_{m}(k r)\sin(\delta_{m}(r)) \right],
\end{equation}
where $A_{m}(r)$ is called the amplitude function and the phase function $\delta_{m}$ is the phase shift arising from a potential cut-off at a distance $r$. To completely define these newly introduced functions $A_{m}(r)$ and $\delta_{m}$ we make the following ansatz for the first derivative of $\chi_{A}(r)$ with respect to $r$
\begin{equation}
\label{derivansatz}
 \chi_{A}'(r) = A_{m}(r)\left[ J_{m}'(k r)\cos(\delta_{m}(r)) - N_{m}'(k r)\sin(\delta_{m}(r)) \right],
\end{equation}
where $'$ denotes differentiation with respect to $r$. Now, setting the direct derivative of Eq.~\eqref{wavefuncansatz} equal to Eq.~\eqref{derivansatz} suggests the following useful condition
\begin{equation}
\label{condition}
 \frac{A_{m}'(r)}{A_{m}(r)} = \delta_{m}'(r) \frac{J_{m}(k r)\sin(\delta_{m}(r)) + N_{m}(k r)\cos(\delta_{m}(r))}{J_{m}(k r)\cos(\delta_{m}(r)) - N_{m}(k r)\sin(\delta_{m}(r))}.
\end{equation}
Upon substituting Eq.~\eqref{wavefuncansatz} and Eq.~\eqref{derivansatz} into the lower coupled Eq.~\eqref{coupledlower} we naturally find the lower radial wavefunction component $\chi_{B}(r)$ is
\begin{equation}
\label{lowerwavefunc}
 \chi_{B}(r) = \frac{A_{m}(r)}{k-V(r)}\left[ \left(-J_{m}'(k r) + \frac{m}{r}J_{m}(k r)\right)\cos(\delta_{m}(r)) - \left(-N_{m}'(k r) + \frac{m}{r}N_{m}(k r)\right)\sin(\delta_{m}(r)) \right].
\end{equation}
We can now utilize the upper coupled Eq.~\eqref{coupledupper}: upon substituting in Eq.~\eqref{wavefuncansatz} and Eq.~\eqref{lowerwavefunc} and eliminating the amplitude function $A_{m}(r)$ via the application of the condition Eq.~\eqref{condition}, we obtain the following first-order differential equation
\begin{align}
\label{phaseeq}
	\begin{split}
	\frac{d}{dr}\delta_{m}(r) = \frac{\pi r}{2} p(r) \left[ \frac{1}{k-V(r)}\frac{dV(r)}{dr} \left( q(r) - \frac{m}{r}p(r) \right)  + \left( V(r)^2 -2 k V(r) \right) p(r) \right], \\
 p(r)= J_{m}(k r)\cos(\delta_{m}(r)) - N_{m}(k r)\sin(\delta_{m}(r)), \\
 q(r)= J_{m}'(k r)\cos(\delta_{m}(r)) - N_{m}'(k r)\sin(\delta_{m}(r)), 
	 \end{split}
\end{align}
where we have introduced the auxiliary functions $p(r)$ and $q(r)$ and in addition have used the Wronskian of the Bessel functions $W\{J_{m}(x), N_{m}(x)\} = J_{m}(x) N_{m}'(x) - N_{m}(x) J_{m}'(x) = -2 / (\pi x)  $ to simplify the final expression.\cite{Abramowitz} Eq.~\eqref{phaseeq} is the so-called \textit{phase equation}, and is subject to the initial condition $\delta_{m}(0) = 0 $, as follows from being in the free particle limit. We can see from Eq.~\eqref{phaseeq} how the potential $V(r)$ gradually accumulates the desired phase shift starting from $\delta_{m}(0) = 0 $ and finishing with the total phase shift of the scattering problem, given by
\begin{equation}
\label{phaseshift}
 \delta_{m} = \lim_{r \to \infty} \delta_{m}(r).
\end{equation}
This condition ensures the phase shift is uniquely defined, avoiding an ambiguity of $\pi$ that appears in other methods.\cite{Newton} When investigating bound states in the massless case we can only consider zero-energy states, where the Neumann function is divergent and so the following condition is implied
\begin{equation}
\label{boundcond2}
	\delta_{m} = n \pi, \qquad n = 1, 2, 3...
\end{equation}
Eq.~\eqref{boundcond2} is related to the Levinson's theorem for massless 2D Dirac particles, which states a relation between the phase shift at zero-momentum and the number of bound states.

Please note when considering the massive 2D Dirac particles, e.g. as found in h-BN or gapped graphene,\cite{Zhou} which allow bound states at finite energy, an equation analogous to the phase equation Eq.~\eqref{phaseeq} can be derived for treating confined states
\begin{align}
\label{phaseeq2}
	\begin{split}
	\frac{d}{dr}\eta_{m}(r) = -r f(r) \left[ \frac{1}{k-V(r)}\frac{dV(r)}{dr} \left(g(r) - \frac{m}{r}f(r) \right)  + \left( V(r)^2 -2 k V(r) \right) f(r) \right], \\
 f(r)= I_{m}(\kappa r)\cos(\eta_{m}(r)) - K_{m}(\kappa r)\sin(\eta_{m}(r)), \\
 g(r)= I_{m}'(\kappa r)\cos(\eta_{m}(r)) - K_{m}'(\kappa r)\sin(\eta_{m}(r)), 
	 \end{split}
\end{align}
where $I_{m}(\kappa r)$ and $K_{m}(\kappa r)$ are the modified Bessel and Neumann functions respectively and the effective wavevector $\kappa=(\Delta^2-k^2)^{1/2}$. Eq.~\eqref{phaseeq2} has been simplified\cite{Abramowitz} by noting the Wronskian of the modified Bessel functions $W\{I_{m}(x), K_{m}(x)\} = I_{m}(x) K_{m}'(x) - K_{m}(x) I_{m}'(x) = -1/x  $. Notably, Eq.~\eqref{phaseeq2} is also relevant for considerations of the surface states on 3-D topological insulators such as $\text{Bi}_2\text{Te}_3$, where in this case the mass term arises from the exchange energy from a magnetic insulator.\cite{Beenakker} When considering bound states we note the modified Bessel function of the first kind is divergent and so the following condition is implied
\begin{equation}
\label{boundcond}
	\eta_{m} = \left( n-\frac{1}{2}\right) \pi, \qquad n = 1, 2, 3 ...
\end{equation}
which is a representation of the Levinson theorem for the massive 2D Dirac equation.\cite{Lin, Dong}

\section{\label{results}Results}

\begin{figure}[htbp] 
\centering
\includegraphics[width=0.5\textwidth]{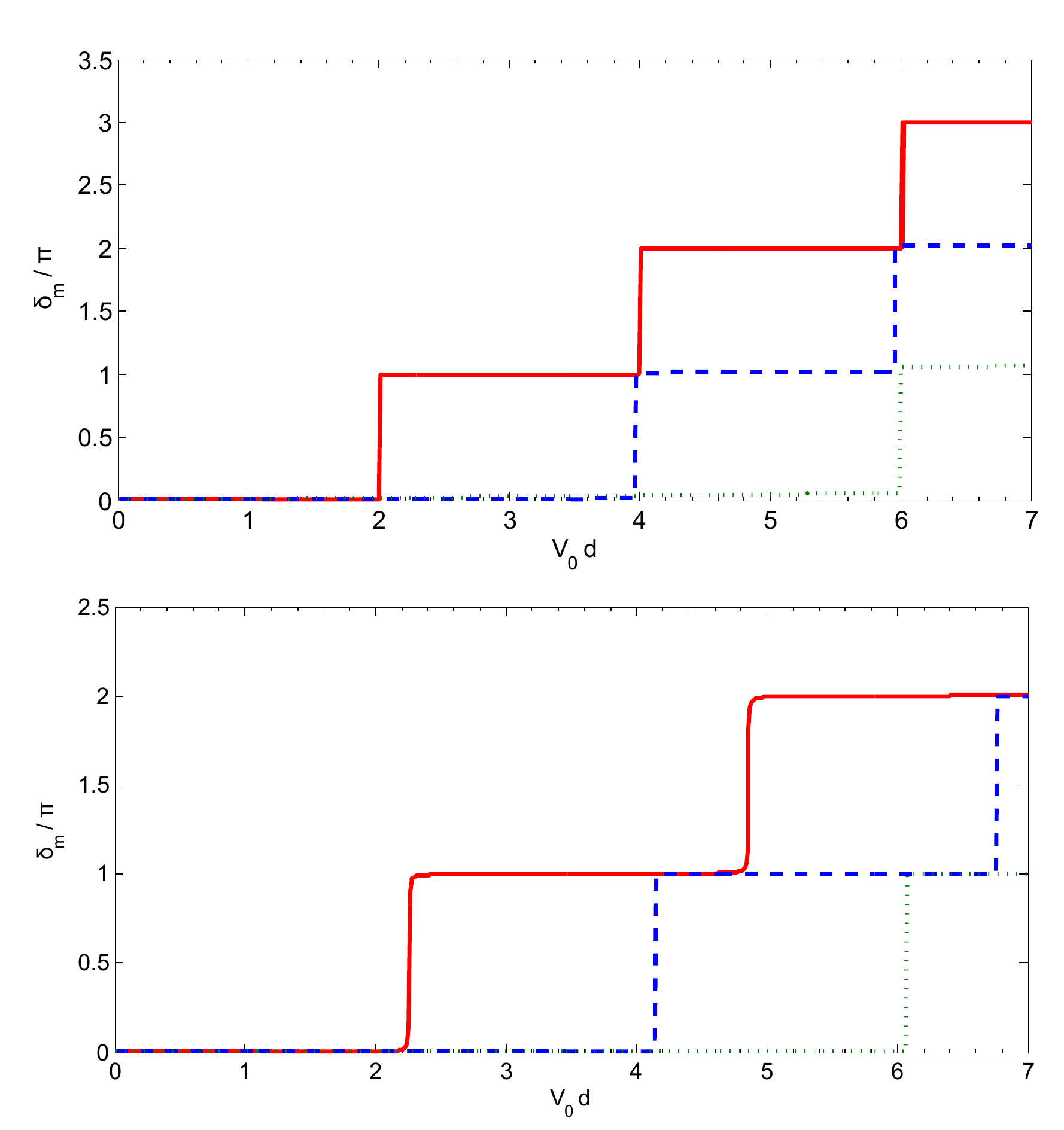}  
\caption{(Color online) Plots of scaled phase shifts $\delta_{m}/\pi$ against potential strength $V_0d$ for massless Dirac particles, of energy tending towards zero, incident on (top) the Lorentzian potential and (bottom) the model potential of Eq.~\eqref{newpotential} with the realistic $1/r^3$ decay. We show results for angular momentum $m=0,1,2$ corresponding to the solid line (red), dashed line (blue) and dotted line (green) respectively.}
\label{fig:fig1}
\end{figure}

We now check that we can reproduce known results by solving the phase equation Eq.~\eqref{phaseeq}, describing the massless charge carriers of graphene, for the case of zero-energy states formed in a Lorentzian potential $V(r)=-V_0/(1+(r/d)^2)$ , an analytically solved problem.\cite{Downing} In this case the condition for bound states when $m\ge0$ is $V_0d=2(N+m)$, where $N$ is a positive integer. Thus, when solving Eq.~\eqref{phaseeq}, the phase equation for massless Dirac particles, when $k\to0$ we should see the threshold value of $V_0d$ reached before step-like behavior as $V_0d$ is turned up and more confined zero-energy states appear. This is exactly what we find in Fig.~\ref{fig:fig1} (top). We also note that the sign of $V_0$ is irrelevant for the creation of zero-energy confined states since a well for an electron is a hill for a hole and vice versa, thus Fig.~\ref{fig:fig1} has mirror symmetry about the $V_0d = 0$ axis. At the Dirac point, when the density of states vanishes yet the conductivity remains finite, these zero-energy states should be important as we shall see in the later on. 

Whilst screening only effects the strength of the Coulomb potential and not its characteristic decay,\cite{DasSarma} due to the quasi-relativistic nature of the carriers in graphene, a cut-off is necessary at the origin and the presence of an image charge in the metallic back gate will lead to a dipole-like $1/r^3$ decay at large distances, thus a convenient choice of model potential is
\begin{equation}
\label{newpotential}
	V(r)=\frac{-V_0}{1+(r/d)^3}.
\end{equation}
In Fig.~\ref{fig:fig1} (bottom) we investigate confined states with Eq.~\eqref{newpotential} and again see a characteristic threshold potential strength product spread $V_0d$, followed by the signature staircase behavior of confined zero-energy states. Of course, compared to the exactly-solvable Lorentzian potential with its accidental degeneracies, the staircase does not share the same beautiful symmetries and the condition for full-confined states can be approximated by $V_0d\approx 2.63 (N+ 0.72 m - 0.13)$. However we can now predict in realistic graphene flakes, where we would expect charged impurities to cause potentials similar to the type  Eq.~\eqref{newpotential}, that there is the possibility for the appearance of fully-confined zero-energy states. 

Note, as we show in Appendix~\ref{appendA} by considering a general class of potential wells regular at the origin and decaying like some power law (which is faster than the Coulomb potential), the toy model interaction potential one chooses is not crucial for  the actual existence of zero-energy bound states, but is important for knowing where one should search for them in terms of potential strength. Thus we have shown that it is possible to create electrostatic traps holding electrons (or holes) in the regime where the Fermi level is close to the Dirac point energy. Then it should be possible to release these trapped quasiparticles into the system by a small adjustment of the trapping potential strength, which will result in the sought-after on/off behavior.

Experimentally, such states should be able to be detected by SPM experiments, as proposed in Ref.~[\onlinecite{Downing}], where smoothly changing either the charge on the SPM tips or their distance above the graphene plane, and continuously holding the Fermi level at the Dirac point using the back-gate should be sufficient to see confined zero-modes. A network of sparse SPM tips, of radius $R_{tip}$ and separated in a square grid defined by an inter-tip separation $s$, all held at a distance $h_2$ above a metallic back gate and $h_2-h_1$ above the graphene plane gives a similarly behaved potential to Eq.~\eqref{newpotential}, but in a more complicated form due to the method of images,
\begin{equation}
\label{sumpot}
	U(r) = \frac{e Q_{tip}}{4 \pi \epsilon_0 \epsilon_r} f(r), 
	\quad f(r) = \sum_{j, k = -n_1}^{n_2} \left( (x+js)^2 + (y+ks)^2 + (h_2-h_1)^2 \right)^{-1/2} - \left( (x+js)^2 + (y+ks)^2 + (h_2+h_1)^2 \right)^{-1/2}.
\end{equation}
Fitting Eq.~\eqref{newpotential} to Eq.~\eqref{sumpot} by matching the functions at both their maximum and half-maximum values, one finds that the tip voltage $V_{tip}$ at which one would see such zero-modes is 
\begin{equation}
\label{newpotential2}
	V_{tip} = \frac{(V_0 d)_{N, m}}{f(0) d} \frac{\hbar v_F}{e R_{tip}}, \quad \text{subject to the constraint}~2 f(d) = f(0),
\end{equation}
where we have seen from Fig.~\ref{fig:fig1} (bottom) the dimensionless parameter $(V_0 d)_{N, m \ne 0} = 4.16, 6.06, 6.79 ...$ such that we are dealing with\cite{numbers} tip voltages of the order of tens of mV, $V_{tip} = 20  \text{mV}, 29  \text{mV}, 33  \text{mV}$ and so on. The existence of confined states opens up the possibility of Coulomb blockade-type physics in graphene. Indeed, the quantum dots created with careful adjustment of the key parameters can be be seen to be `optimal traps'. An estimate of the charging energy of the optimal trap, using a simple disc capacitor model, shows a charging energy of the order of $\text{meV}$, thus such effects could be seen at room temperature. The effect arises due to the tightly confined nature of the wavefunction in optimal traps which leads to small capacitance and a significant charging energy, which is negligible for the usual deconfined states. 
  
\begin{figure}[htbp] 
\includegraphics[width=0.8\textwidth]{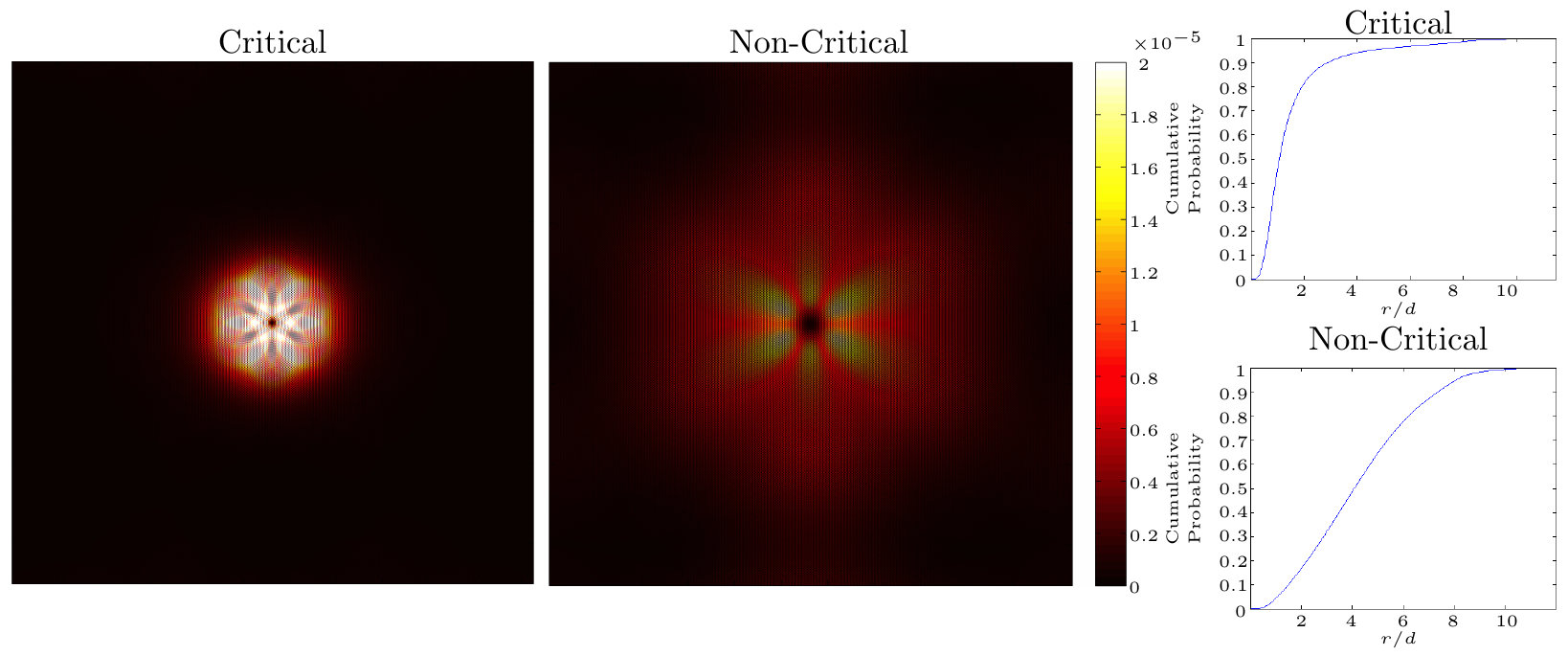}  
\caption{(Color online) Probability density plots of near zero-energy states confined within the model confining potential Eq.~\eqref{newpotential}, as calculated via tight-binding methods. We show example critical $V_0 d = 4.10$ (left) and non-critical $V_0 d = 2.60$ (right) states, as well as the associated cumulative probability plots (right). }
\label{fig:fig2}
\end{figure}

We also show in Fig.~\ref{fig:fig2} that the zero-modes of the Dirac equation in the potential Eq.~\eqref{newpotential}, as predicted in the continuum model, are indeed present as shown via tight-binding calculations.\cite{Zarenia, Chaves} It is found the wavefunctions have a ring-like structure, which ensures avoidance of any Klein tunneling effects, i.e the states are zero-energy vortices with $m \ne 0$. It is most noticeable how by adjusting the parameter $V_0 d$ one can go from a tightly confined state to a state with a highly spread probability density. We have also checked the zero-modes are indeed robust to the shape of the graphene flake.

As mentioned previously, once the scattering phase shift is known a number of other useful physical properties can be quickly calculated. The partial cross-section $\zeta_m$  can be taken as
\begin{equation}
\label{partialcrosssection}
 k \zeta_m = \sin^2(\delta_m),
\end{equation}
whilst the total scattering cross-section $\zeta$, a measure of the length felt by oncoming particles, and the transport cross-section $\zeta_{T}$ easily follow from the scattering phase shifts via
\begin{equation}
\label{crosssection}
 \zeta = \frac{4}{k} \sum\limits_{m=-\infty}^\infty \sin^2(\delta_m), \quad \zeta_{T} = \frac{2}{k} \sum\limits_{m=-\infty}^\infty \sin^2(\delta_{m+1}-\delta_{m}),
\end{equation}
and we note for low-energy scattering we can take the $s$-wave approximation, i.e. only small $m$ need to be considered as Eq.~\eqref{crosssection} is derived from partial-wave expansions where terms with high $m$ are negligible when $k \to 0$. Detailed derivations of Eqs.~(\ref{partialcrosssection},\,\ref{crosssection}) via 2D elastic scattering theory adapted for Dirac fermions have been given by Novikov in Ref.~[\onlinecite{Novikov}].

\begin{figure}[htbp] 
\includegraphics[width=0.8\textwidth]{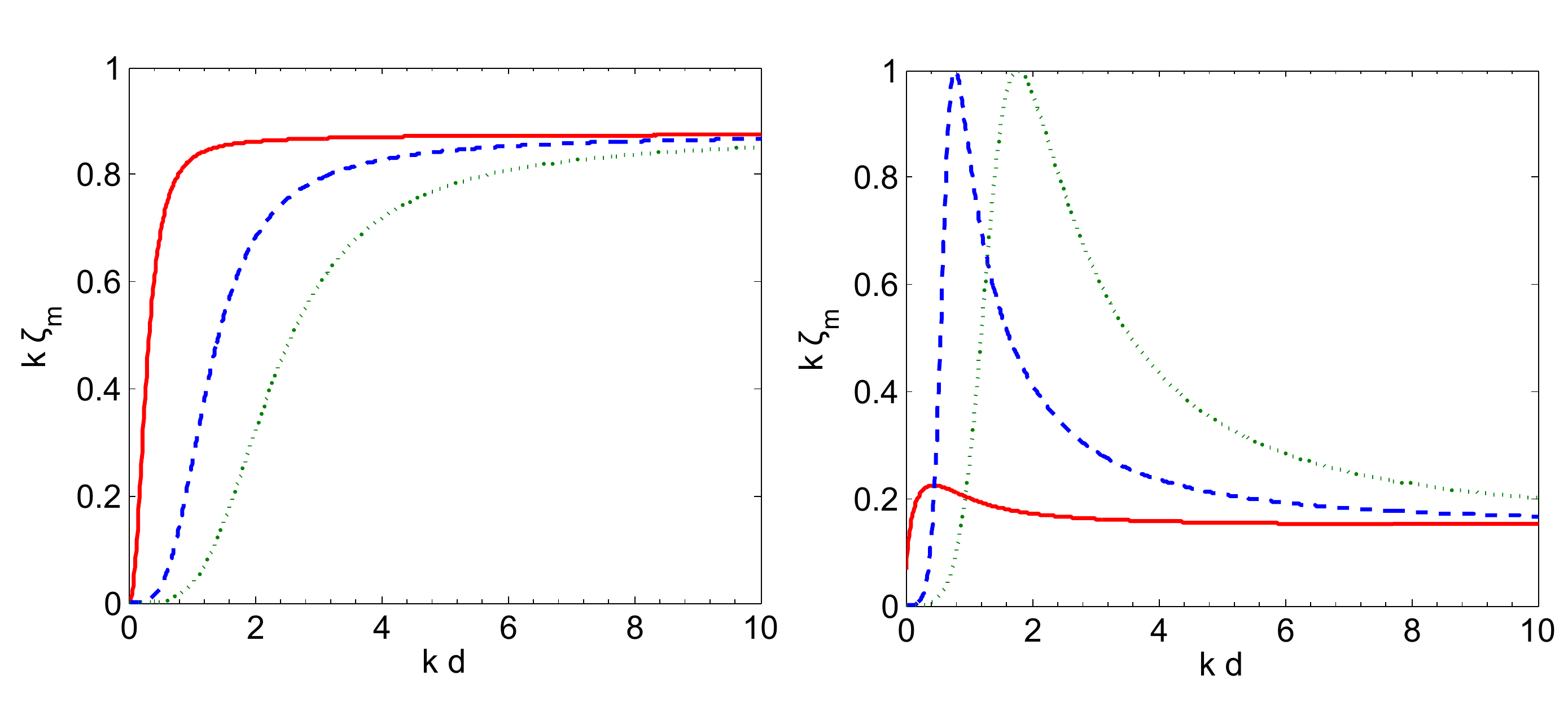}  
\caption{(Color online) A plot of dimensionless partial scattering cross-section $k \zeta_m$ as a function of scaled energy $k$ for massless Dirac particles incident on the model potential with $1/r^3$ decay given by Eq.~\eqref{newpotential2}, with example noncritical potential strength (left) $V_0 d =1.00$ and critical (right) $V_0 d =2.27$. We show results for $m=0,1,2$, corresponding to the solid line (red), dashed line (blue) and dotted line (green) respectively.}
\label{fig:fig3}
\end{figure}

The energy dependence of scattering cross-section has has previously been considered for a square well\cite{Chaplik, AndreGeim} and we outline a thorough solution to this problem in Appendix~\ref{appendB}. We will now revisit the problem with a smooth, short-range potential given by Eq.~\eqref{newpotential}, which is relevant for gated structures or for hypercritical charges. We show in Fig.~\ref{fig:fig3} plots of dimensionless partial cross-section $k \zeta_m$ for the case of non-critical (left) and critical (right) parameters of the potential Eq.~\eqref{newpotential}. We find, in contrast to the general case, at the critical potential strength the $m=0$ non-square-integrable state goes not go to zero in $k \zeta_m$ as quickly as $k$ tends towards zero, and thus is divergent in partial cross-section $\zeta_m$ as energy tends towards zero. This is because $m=0$ is a resonant state with a non-normalizable wavefunction (and a non-zero probability density at the origin $r=0$) and so has such a particle has an enhanced likelihood of being scattered. Of course, this behavior remains in calculations of transport scattering cross-section via Eq.~\eqref{crosssection}.

We also note, in striking contrast to non-relativistic particles, at large energies the phase shift $\delta_{\infty}$ is non-zero.\cite{Parzen} Remarkably, $\delta_{\infty}$ is also angular momentum-independent and we find from Eq.~\eqref{phaseeq} the explicit form
\begin{equation}
\label{large}
 \delta_{\infty} = - \int_0^{\infty} V(r) \mathrm{d}r,
\end{equation}
such that for the considered potential Eq.~\eqref{newpotential2} the dimensionless partial scattering cross-section in the large energy limit is given by $k \zeta_m \to \sin^{2}(\frac{2 \pi}{3 \sqrt{3}} V_{0} d)$, as displayed in Fig.~\ref{fig:fig3}. Thus, the transport cross-section $\zeta_{T}$ vanishes in this limit.

\begin{figure}[htbp] 
\includegraphics[width=0.4\textwidth]{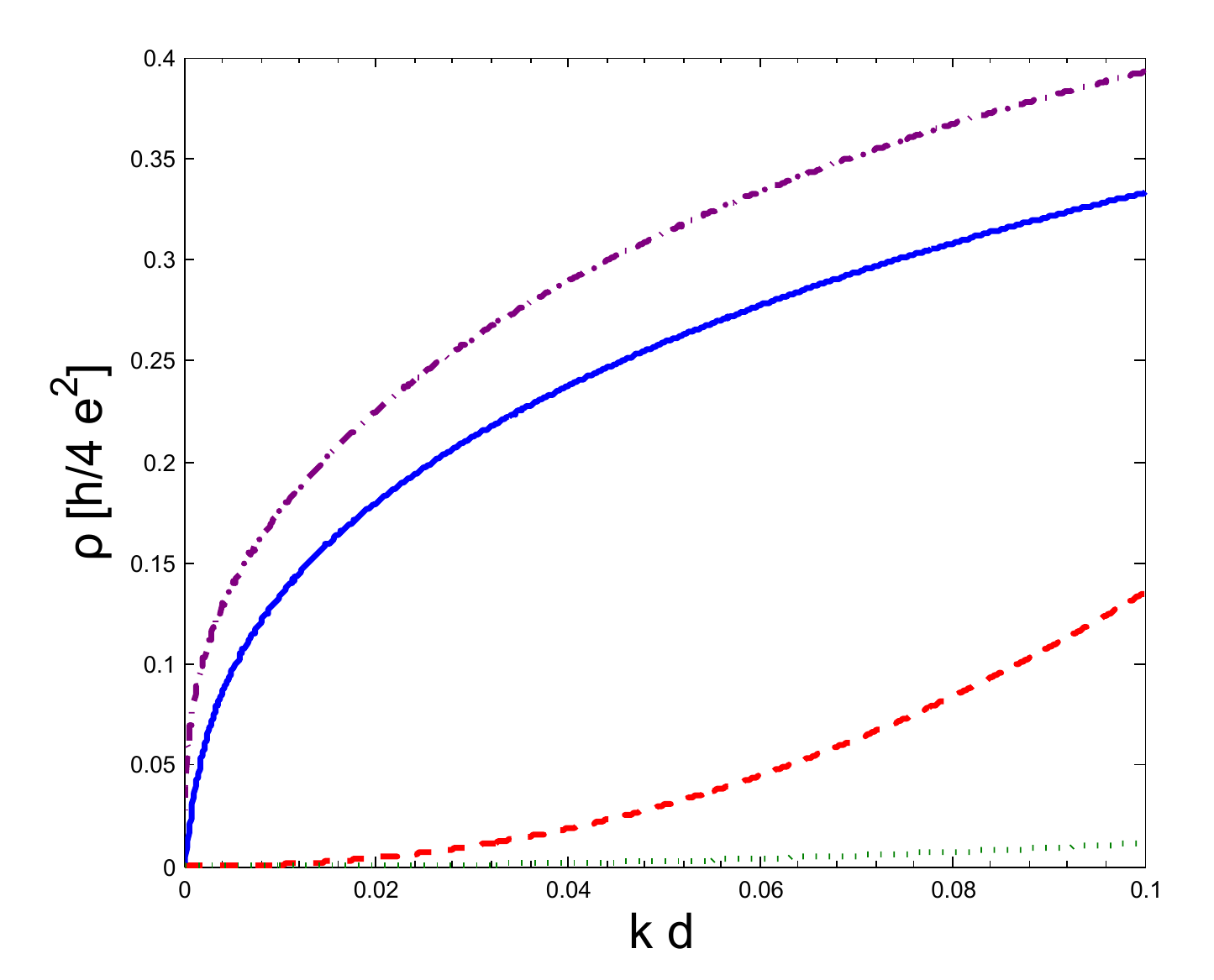}  
\caption{(Color online) A plot of resistivity $\rho$, measured in units of $h/4 e^2$, as a function of energy $k d$ for massless Dirac particles scattering on the model potential with $1/r^3$ decay. We show results for example uncritical cases $V_0 d =1.00, 3.00$, corresponding to the dashed line (red) and dotted line (green) respectively, and example critical cases $V_0 d = 2.27, 4.87, ...$, corresponding to the solid line (blue) and dash-dot line (purple) respectively. We set $2 n_s = \pi n_e$ to more clearly contrast the differences due to the effect of cross-section only.}
\label{fig:fig4}
\end{figure}

The smoking gun of confined zero-modes in the laboratory could be via their contribution to resistivity, which in semiclassical Boltzmann theory can be expressed as\cite{AndreGeim}
\begin{equation}
\label{resistivity}
 \rho = \frac{h}{4 e^2} \frac{2 n_s}{\pi n_e} k \zeta_T
\end{equation}
where $n_s$ is the density of scatterers and the electron density $n_e = k^2 / \pi$. We show in Fig.~\ref{fig:fig4} the behavior of resistivity at small $k d$. The presence of confined zero-modes sees a much slower drop to zero than in the critical cases, such that when one factors in the energy dependence of $n_e \propto k^2$, there is a divergence in resistivity contribution at energies tending towards zero in the critical case only, $\rho [h/4 e^2] \to \infty$. It should be possible to utilize this consequent drastic suppression of mobility to create an off-state in graphene, which can be explored artificially with a series of point gates held above the graphene monolayer. In the non-resonant case,  when taking into account $n_e \propto k^2$, we find as $k \to 0$ resistivity tends to a constant (depending on the number of scatterers $n_s$) as expected.

\section{\label{disc}Discussion}

We have derived the phase equations for the 2-D Dirac equation using the VPM, suitable for use straightaway in scattering calculations concerning Dirac materials such as graphene. In doing so, we provide a numerical (experimental) proof of the Levinson Theorem for massless 2-D particles. These phase equations are of first-order, and so relatively undemanding computationally, and have solutions in terms of scattering phase shifts, thus other desired scattering properties readily follow. Applying the method to fully-confined states in graphene, we reproduce an exact result from the literature and go on to investigate a more physical potential again finding that a certain potential strengths and spreads zero-energy bound states are likely to form, which is most important when describing resonant scattering. 

We have also calculated the energy dependence of scattering cross-section, finding a major distinction for the $m=0$ mode for critical potential strengths (those able to support truly bound states). In this special case, we predict a dominant peak in scattering cross-section at zero-energy, suggesting a high probability of being scattered, which can be explored experimentally via scanning probe microscopy. In an experiment with a series of point gates above the graphene layer, one may be able to use artificial resonant scattering to switch off the chiral tunneling effect found in monolayer graphene by greatly reducing the mobility of carriers.

It should also be mentioned that the existing experiments on atomic clusters on a graphene surface can arguably be explained by the considered zero-energy states without involving the atomic collapse picture. These experiments deal with back-gated graphene which precludes the use of a long-range Coulomb potential. The measured wavefunction density has a pronounced ring-like structure on a scale of many lattice constants, instead of a sharp peak at the atomic scale as expected from `fall-into-the-center' physics. Moreover, the observed features in the density of states are very close to zero energy (in fact, on both sides of it) instead of being a few electron volts away from this energy as expected for a collapsed state.

\section*{Acknowledgments}
This work was supported by the EPSRC, the EU FP7 ITN NOTEDEV (Grant No. FP7-607521), and FP7 IRSES projects CANTOR (Grant No. FP7-612285), QOCaN (Grant No. FP7-316432), and InterNoM (Grant No. FP7-612624). During this work, CAD was supported by a swivel chair with footrest.

\begin{appendix}

\section{\label{appendA}Bound states in potential wells decaying as a power law}

Let us examine the condition on bound states to form in regularized potential wells decaying asymptotically as a power law governed by a parameter $p>1$, namely
\begin{equation}
\label{power1}
	V(r) =  -V_{0}
  \begin{cases}
   1, & \text{if } r \le R \\
   \left( \tfrac{R}{r} \right)^p ,  & \text{if } r > R
  \end{cases}
\end{equation}
such that the wavefunction components inside the well are simply Bessel functions and outside the well are Bessel functions, in a new variable $\xi = \tfrac{V_0 R}{p-1}\left(\tfrac{R}{r}\right)^{p-1}$, product the function $r^{-p/2}$ . Bound zero-energy states can never arise for $m=0, -1$, due to the requirement of some rotation to combat the Klein tunneling phenomenon.\cite{Downing} Furthermore, some other negative angular momentum states $m_e \le m \le -1$ are excluded depending on the power of the decay, for $p > -2 m_e$. Thus, as one considers increasingly short-range interactions, one will find a wider band of `missing' nonpositive integer angular momentum states cumulating in $m_e$. The eigenenergies corresponding to the square-integrable solutions can be found from the determinant of the matrix
\begin{equation}
\label{power2}
\begin{vmatrix}
 J_{|m|}( V_0 R) \quad J_{|m+1|}( V_0 R) 
		\\  J_{|\alpha|}( \tfrac{V_0 R}{p-1}) \quad J_{|\alpha-1|}( \tfrac{V_0 R}{p-1}) 
\end{vmatrix}
		=0, \quad \alpha = \frac{m+p/2}{p-1}
\end{equation}
which can be solved by the usual root-finding methods for the values of $V_0 R$ which are able to support confined modes. In the limit of large $V_0 R > > 1$, Eq.~\eqref{power2} can be rewritten in terms of elementary functions
\begin{equation}
\label{power3}
\begin{vmatrix}
 \cos \left( V_0 R - \tfrac{\pi}{2} |m|  - \tfrac{\pi}{4} \right) \quad \cos \left( V_0 R - \tfrac{\pi}{2} |m+1|  - \tfrac{\pi}{4} \right) 
		\\  \cos \left( \tfrac{V_0 R}{p-1} - \tfrac{\pi}{2} |\alpha-1|  - \tfrac{\pi}{4} \right) \quad \cos \left( \tfrac{V_0 R}{p-1} - \tfrac{\pi}{2} |\alpha|  - \tfrac{\pi}{4} \right) 
\end{vmatrix}
		=0, 
\end{equation}
and the most symmetric case $p=2$ is equivalent to $V_0 R = \tfrac{\pi}{4} \left( |m| + |m+1| + 2n +1\right)$, where $n$ is an integer.

Whilst this somewhat kinky class of potential is not as realistic as the smooth potentials considered in the main body of this work, it quantitatively demonstrates how bound zero-energy states are realizable in many situations for fast-decaying potentials. As expected, a separate analysis for the regularized Coulomb potential ($p=1$) yields no possibility for bound zero-energy states. This conclusion is formed as the associated transcendental formed can never have a solution.

\section{\label{appendB}Scattering by a circular finite potential well}

\begin{figure}[htbp] 
\includegraphics[width=0.7\textwidth]{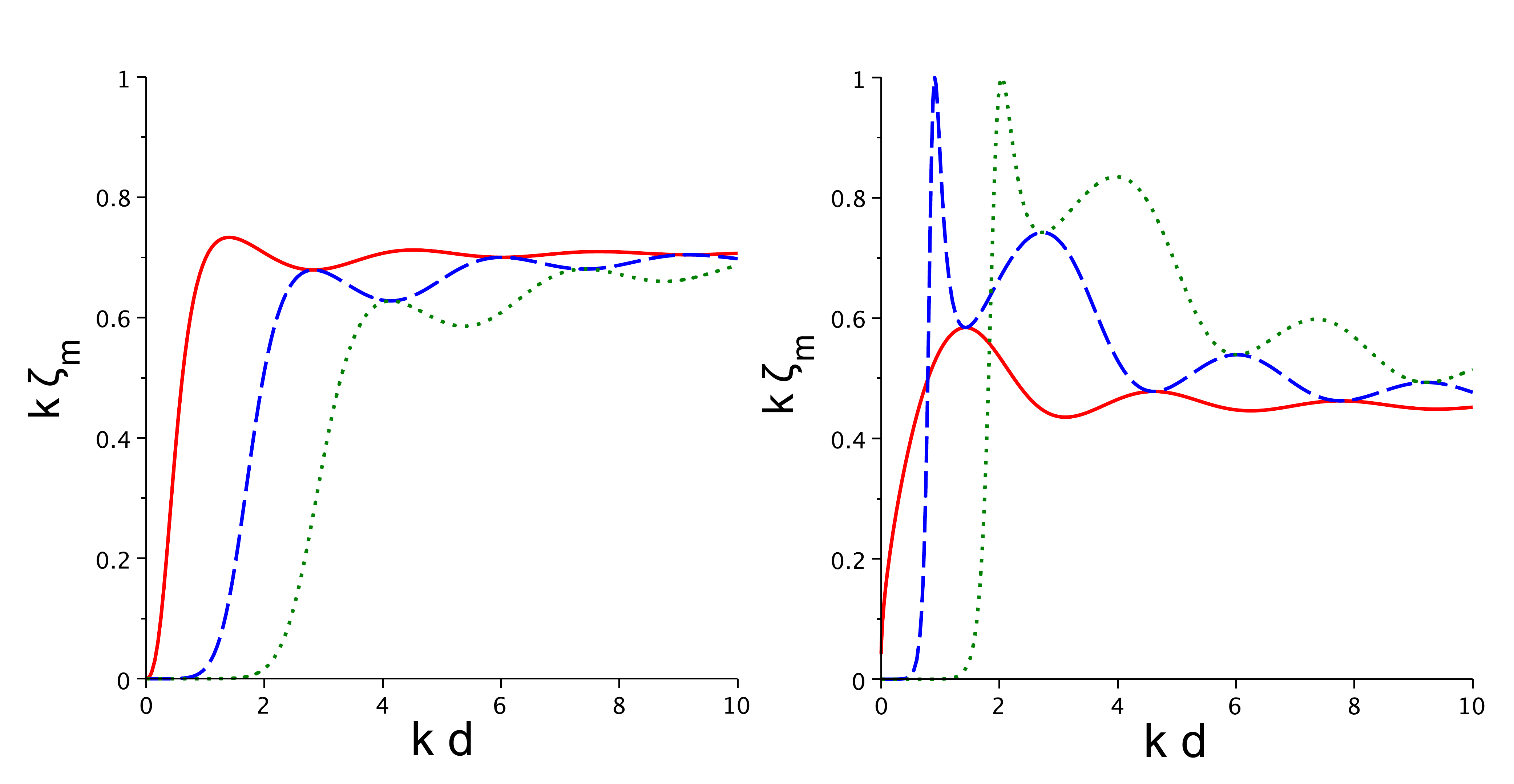}  
\caption{(Color online) A plot of dimensionless partial scattering cross-section $k \zeta_m$ as a function of scaled energy $k$ for massless Dirac particles incident on the circular finite potential well, with noncritical potential strength (left) $V_0 d =1$ and critical (right) $V_0 d =2.40$. We show results for $m=0,1,2$, corresponding to the solid line (red), dashed line (blue) and dotted line (green) respectively.}
\label{fig:fig5}
\end{figure}

Let us consider the example of a circular finite potential well\cite{Chaplik} given by $V(r) = -V_{0} \Theta(d-r)$, where $\Theta (z)$ is the Heaviside step function, which allows us to obtain the following solutions of  Eq.~\eqref{decoupledchi}: $\chi_{A}(r) = A_{m} J_{m}((k + V_{0}) r)$ inside the well and $\chi_{A}(r) = B_{m} \left[ J_{m}(k r) \cos(\delta_{m}) - N_{m}(k r) \sin(\delta_{m}) \right] $ outside the well. Upon finding $\chi_{B}(r)$ from Eq.~\eqref{coupledlower} and matching both wavefunction components at the boundary $r = d$, we obtain the following expression for the tangent of the phase shift
\begin{equation}
\label{tangentphase}
	\tan(\delta_{m}) = \frac{ J_{m}((k + V_{0}) d)~J_{m+1}(k d) - J_{m+1}((k + V_{0}) d)~J_{m}(k d) } 
	{ J_{m}((k + V_{0}) d)~N_{m+1}(k d) - J_{m+1}((k + V_{0}) d)~N_{m}(k d) }. 
\end{equation}
It can be seen that resonances in scattering cross-section do exist at certain energies, as can be seen by substituting Eq.~\eqref{tangentphase} into Eq.~\eqref{partialcrosssection}, or into the more convenient form $k \zeta_m = 1/(1+ \tan(\delta_m)^{-2})$. At energies tending to the Dirac point energy $k \zeta_m$ and, more importantly, the true partial cross-section $\zeta_m$ goes to zero. At large energies we find the expected constant value, given by $k \zeta_m \to \sin^{2}(V_{0} a)$, as follows from Eq.~\eqref{large}. Both of these features can be seen in Fig.~\ref{fig:fig5} (left). 

One can find the most prominent resonances of partial cross-section for this problem, defined by $k \zeta_m =1$, which can only occur for some specific potential strengths, which satisfy the condition $J_{m}((k + V_{0}) d) N_{m+1}(k d) = J_{m+1}((k + V_{0}) d) N_{m}(k d) $. If there is a solution to this equation, it is straight forward to find the resonance energy, which is the global maximum solution of $\frac{d}{d k} \zeta_m = 0$. 

In the s-wave approximation, one can see by using Eq.~\eqref{tangentphase} in conjunction with Eq.~\eqref{crosssection} the contribution to resistivity, calculated via Eq.~\eqref{resistivity}, is given by\cite{AndreGeim}
\begin{equation}
\label{resis1}
	\rho \approx \frac{h}{4 e^2} n_s d^2 
\end{equation}
which, of course, can be neglected when the concentration of scatterers and/or the scattering radius $d$ is small. However, this analysis neglects the possible existence of confined zero-modes, which can be determined by the condition $J_{m}(V_{0} d) = 0$. The effect of the extended $m=0$ state (which occurs at the zeroes of the zeroth Bessel function, $V_{0} d = 2.40, 5.52, ... $) is to see the quantity $\tan(\delta_{m})$ approach zero slowly (logarithmically) as we approach the Dirac point energy, such that $k \zeta_0$ effectively tends to a constant, producing a `super-resonance' at zero-energy in pure partial cross-section $\zeta_0$, as displayed in Fig.~\ref{fig:fig5} (right).  This leads to a noticeable contribution to resistivity\cite{AndreGeim}
\begin{equation}
\label{resis2}
	\rho \approx \frac{h}{4 e^2} \frac{n_s}{n_e} \frac{1}{\ln^2(k d)}  
\end{equation}
which is clearly more dominant than that found in Eq.~\eqref{resis1} in the s-wave approximation.

\end{appendix}


\begin{thebibliography}{100}
\bibitem{note}This condition prohibits potentials decaying slower than $r^{-2}$ as $r \to \infty$. Otherwise the asymptotic solution of Eq.~\eqref{decoupledchifree} always adds to the phase shift however large $r$ becomes, as is familiar from the asymptotic behavior of the Coulomb wavefunction in the non-relativistic case.
\bibitem{numbers}In this estimate, we use $h_2 = 300 \text{nm}$, $h_1 = 200 \text{nm}$, $s = 40 \text{nm}$, $R = 5 \text{nm}$ and $n_1 = n_2 = 2$.
\end{thebibliography}
\end{document}